\documentclass[%
superscriptaddress,
reprint,
nofootinbib,
nobibnotes,
amsmath,amssymb,
aps,
floatfix,
]{revtex4-1}

\usepackage[utf8]{inputenc}
\usepackage{tikz}
\usetikzlibrary{shapes.geometric}
\usepackage{xcolor}
\usepackage{placeins}
\usepackage{verbatim}
\usepackage{graphicx}
\usepackage{dcolumn}
\usepackage{bm}
\usepackage{hyperref}
\usepackage{caption}
\usepackage{subcaption}
\usepackage{float}
\usepackage{soul}

\begin{document}

\title{Implications of Kleinian relativity}

\author{F. A. P. Alves-J\'{u}nior}
\email{francisco.artur@univasf.edu.br} 
\affiliation{Universidade Federal do Vale do S\~{a}o Francisco, \textit{Campus} Serra da Capivara, Brazil}

\author{A. B. Barreto}
\email{adriano.barreto@caxias.ifrs.edu.br}
\affiliation{Instituto Federal de Educa\c{c}\~{a}o, Ci\^{e}ncia e Tecnologia do Rio Grande do Sul, \textit{Campus} Caxias do Sul, Brazil}
\affiliation{Departamento de F\'{i}sica, Universidade Federal do Paran\'{a}, Curitiba, Brazil}

\author{F. Moraes}
\email{fernando.jsmoraes@ufrpe.br}
\affiliation{Departamento de F\'{i}sica, Universidade Federal Rural de Pernambuco, 52171-900 Recife, PE, Brazil}
\date{\today}
\begin{abstract}
Inspired in metamaterials, we present a covariant mechanics for particles in Kleinian spacetime  and show some of its effects, such as time contraction and length dilatation. We present the new  expressions for relativistic momentum and energy for a point-like particle. To illustrate the new mechanics, we describe the  particle motion under a uniform Newtonian gravitational field. We also  revisit the free  spin-half  particle problem in  Kleinian spacetime, discuss some quantum implications, like the constraint on the dispersion relation for Weyl fermions, and adapt a metamaterial analog system to Klein spacetime.
\end{abstract}
\maketitle
\section{Introduction}
Metamaterials are artificial structures that have peculiar physical properties used in many applications like  electromagnetic wave cloaking, the construction of hyper-lenses, the reproduction of light concentration phenomena, and the reverse Doppler effect \cite{ALITALO200922,smith2004metamaterials, azevedo2018optical,memarian2013light,lee2010reversed}. Furthermore, they provide insights for new analogue models in physics. For instance, Special Relativity, Cosmology, Black Hole Physics, and Celestial Mechanics can be mimicked using metamaterial systems \cite{smolyaninov2015experimental,figueiredo2017cosmology,lu2010simple,genov2009mimicking}. However, one special feature is that metamaterials also make possible the effective realization of the so-called  Kleinian spacetime (KST) 
\cite{smolyaninov2010metric}.

Unlike the Minkowski spacetime, which is associated with a hyperbolic geometry, the Klein spacetime is associated with an ultra-hyperbolic geometry  \cite{lee2006riemannian}. In four dimensions, this spacetime is characterized by the $(+,+,-,-)$ metric signature and  appears in the mathematical theory of ultrahyperbolic differential equations  \cite{john1938ultrahyperbolic}. Such geometry has also appeared in discussions of metric signature change   \cite{alty1994kleinian,alty1996initial}, which is an issue present in classical and quantum cosmological models \cite{sakharov1991cosmological,hartle1983wave,ellis1992covariant}. Moreover, KST is particularly useful in the construction of $\mathcal{N}=2$ super-string models  \cite{barrett1994kleinian,ooguri1990self}.

Examples of ultrahyperbolic geometry are found in theories with more than one time dimensions. This kind of theoretical proposal can lead to closed time-like curves and consequently to issues involving causality. Nevertheless, consideration of this double coordinate time hypothesis is recurrent in the literature. For instance, in Kaluza-Klein models, compactified time-like extra dimensions have been considered to avoid the cosmological constant problem \cite{aref1986extra} and in the Brane World context, the time-like extra dimension could give the \textit{graceful exit} from the de Sitter initial expansion of the universe \cite{maier2013bouncing}. A good general discussion of the number of time-like dimensions and space-like dimensions in a given manifold is presented in \cite{tegmark1997dimensionality} and other interesting theoretical questions, like the realization of two-time  physics,  have been investigated in \cite{bars2020duality,araya2014generalized}.

Motivated by the ``special relativity''  theories constructed in non-Minkowskian, like the de Sitter  \cite{aldrovandi2007sitter,aldrovandi2009sitter} and the anti-de Sitter \cite{sokolowski2016every,sokolowski2016bizarre} spacetimes, and by the fact that metamaterials could allow to simulate some properties of the behavior of particles and fields in KST, we ask ourselves  what could be the relativistic physical implications of the Kleinnian geometry. To investigate mainly the  $x-$direction particle motion in KST, where the Lorentz boost fails, and to give an answer to the above question, we construct a covariant mechanical model and show some of its classical consequences.
Then we apply it in quantum relativistic mechanics to  describe the Weyl fermions  and other features related with one-half spin particles. We also briefly discuss  a metamaterial analogue model for these fermions.

This paper is organized as follows. In section II, we start presenting how the  Kleinian metric appears effectively in metamaterials. In section III, by applying the relativity principles, we find the transformations between the {$xt-$ coordinates} of two inertial reference frames, that we call Klein transformations. In  section IV, we analyze kinematic relativistic effects, such as dilatation of length and contraction of time. In section V, 
we discuss some dynamical aspects. In section VI, we show the behavior of a  particle under an uniform Newtonian gravitational field in KST. In section VII and VIII, we analyse the free and the analogue fermions, respectively. 
In the last section, we close the paper with our final remarks and conclusions.
\section{Metamaterials and Klein spacetimes}

Metamaterials were idealized by Veselago, in 1968 \cite{veselago1968electrodynamics}, which proposed a hypothetical material admitting simultaneously negative permittivity $\epsilon_r$ and permeability $\mu_r$, keeping a real refractive index $n = \pm\sqrt{\epsilon_r \mu_r}$. In the last years this idea has been realized, since  structures have been designed  with negative effective permittivity and/or negative effective permeability, such as a metamaterial composed of copper split ring resonators  or any other left hand metamaterial. Beyond this achievement, other systems, the so-called hyperbolic metamaterials, with  $\epsilon_r<0$ and $\mu_r>0$ or $ \epsilon_r>0$ and $\mu_r<0$ have also been realized using appropriate  technology. 

Considering  anisotropic metamaterials, where the permittivity and permeability are tensors, we find interesting physical possibilities. For example, multilayer
metal-dielectric or metal wire array structures  have been theoretically used to simulate  effective Kleinian spacetimes \cite{smolyaninov2010metric}. In such systems the permittivity tensor $\epsilon_{ij}$ is given by  
\begin{equation}
    \epsilon_{ij}=
    \begin{pmatrix} 
     \epsilon_1 & 0          & 0 \\ 
     0          & \epsilon_1 & 0 \\
     0          & 0          & \epsilon_2\\
    \end{pmatrix},
\end{equation}
where $\epsilon_1>0$ and $\epsilon_2<0$. This tensor works as an effective metric for the electromagnetic wave. As a consequence, an incident electromagnetic wave has its ``extraordinary'' component propagating as a real scalar Klein-Gordon field over a Kleinian spacetime. Another way to simulate KST is to use electronic metamaterials as done in \cite{figueiredo2016modeling}.

{Other examples of anisotropic metamaterial, in our perspective, are the ones based on nematic liquid crystals} \cite{pawlik2014liquid}. In a particular  case \cite{fumeron2015optics}, light rays have been studied  in terms of the effective metric in cylindrical coordinates
\begin{equation}
    ds_{\textrm{eff}}^2=-c^2dt^2+\epsilon_0dr^2 +\epsilon_er^2d\phi^2+\epsilon_0dz^2,
\end{equation}
in a configuration where  the permittivities $\epsilon_0 >0$  and $\epsilon_e <0$, corresponding to a Klein-like spacetime signature.
\section{The spacetime and the Kleinian transformations}
\begin{figure}[H]
    \centering
    \begin{subfigure}[b]{0.45\textwidth}
        \centering
        \includegraphics[scale=0.6,angle=90]{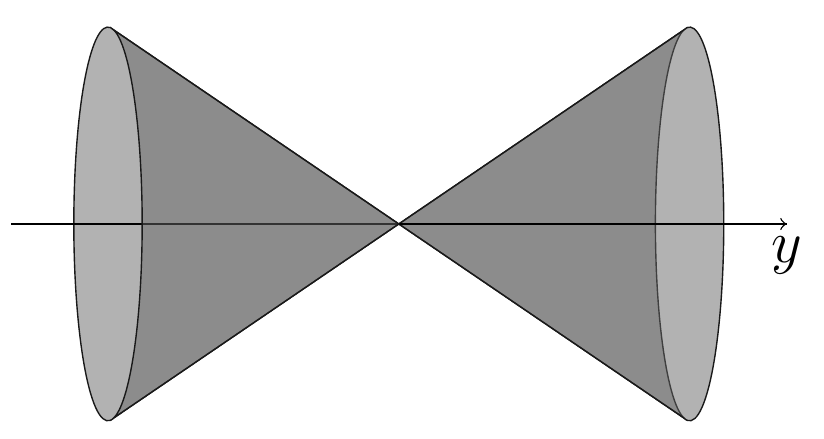}
        \caption{The $(1+2)$ Klein light cone.}
        \label{fig:cone1}
        \end{subfigure}
     \hfill
   \begin{subfigure}[b]{0.45\textwidth}
        \centering
        \includegraphics[scale=0.6]{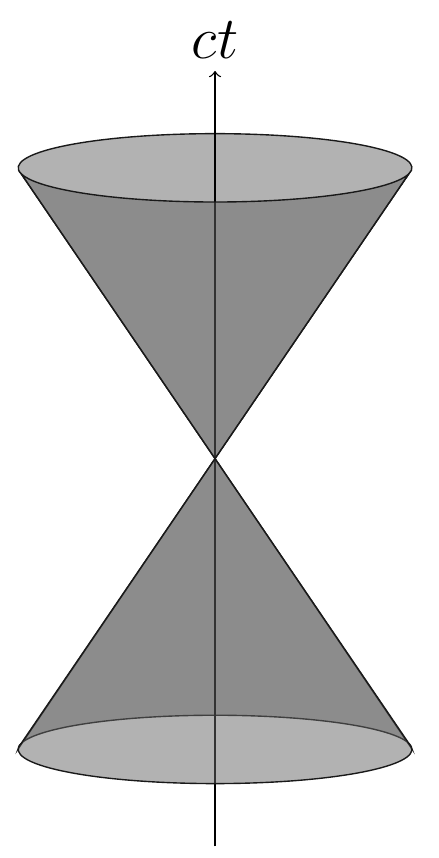}
        \caption{The $(1+2)$ Minkowski light cone.}
        \label{fig:cone2}
        \end{subfigure}
       \caption{Representations of the light cone in the Klein and Minkowski spacetimes.}
\end{figure}

In a more careful way, a Klein metric is a flat pseudo-Riemannian metric, that for simplicity can be stated according to the following line element 
\begin{equation}\label{metrica1}
ds^2=\eta_{\mu\nu}dx^{\mu}dx^{\nu},
\end{equation}
where $\eta_{\mu\nu} = \textrm{diag}(+c^2, +1, -1, -1)$
and $c$ denotes the speed of light in vacuum. From the point of view of Einstein's General Relativity, this metric is a simple mathematical non-Lorentzian solution. For the sake of illustration of the difference between the Lorentzian and Kleinian spacetimes, we show the $(1+2)$ light cones, respectively, in Fig. \ref{fig:cone1} and Fig. \ref{fig:cone2}. 
From these figures we infer that the non-causal regions of one cone are the exact causal regions of the other. In this way, a time-like vector in the Kleinian light cone 
corresponds to a space-like vector outside of the Minkowski light cone, and a space-like vector in Klein cone corresponds to a time-like vector in the other cone.

\begin{figure}[H]
    \centering
    \includegraphics[scale=0.47]{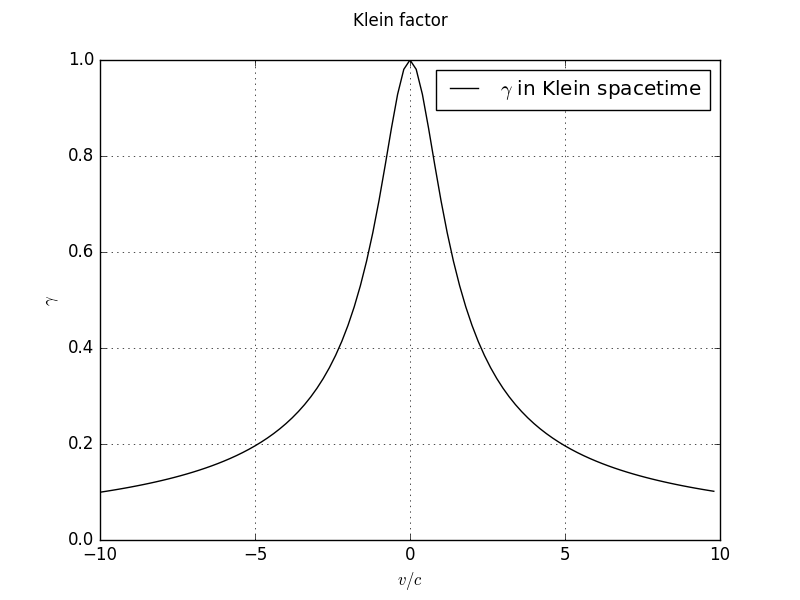}
\caption{The $\gamma$ factor against speed $v/c$. }
    \label{fig:gammabehavior}
\end{figure}

\begin{figure}[H]
    \centering
    \includegraphics[scale=0.47]{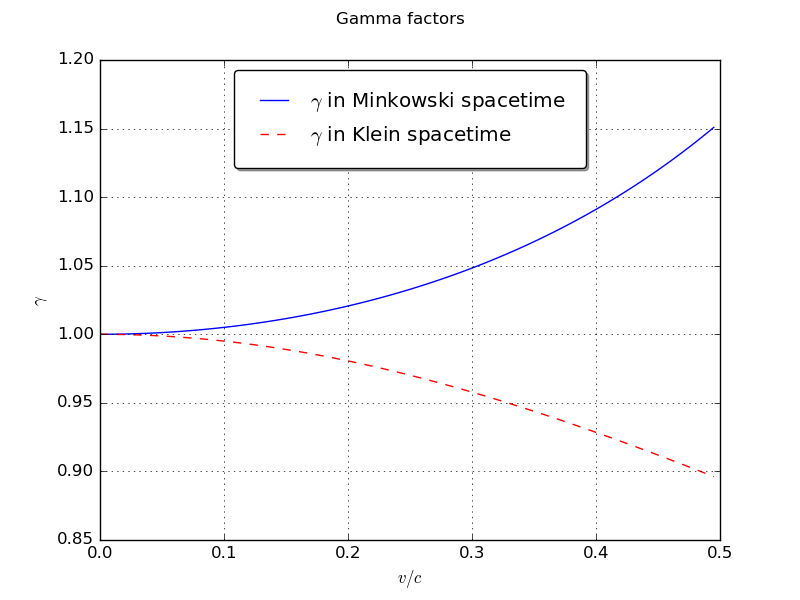}
\caption{Behavior of Lorentz and Klein factors in the two spacetimes}
    \label{fig:gammas}
\end{figure}

Inspired by special relativity theory and, in order to present a covariant model, we postulate that the metric is invariant under a class of  linear transformations, that connects different inertial observers in Kleinian spacetime. Following this assumption, $ds = ds'$, we  deduce a suitable class of coordinate transformations $(t, x, y, z) \rightarrow  (t', x', y' ,z')$, which we call Kleinian transformations.  In particular, taking $y'=y$, $z'=z$, the ``boost'' in $x$ direction is given by
\begin{eqnarray}
x' & = &\frac{x\mp vt}{\sqrt{1+v^2/c^2}}, \label{ktransf1} \\
t' & = &\frac{t\pm \frac{vx}{c^2}}{\sqrt{1+v^2/c^2}} . \label{ktransf2}
\end{eqnarray}
The  Klein factor, $\gamma \doteq \frac{1}{\sqrt{1+v^2/c^2}}$, has a bell-shape profile, as illustrated in the Fig. \ref{fig:gammabehavior}.
\FloatBarrier
In  Fig. \ref{fig:gammas}  we compare the Klein factor with the Lorentz factor profile, and note that for small speed values the two factors are almost the same. However,  for larger values of $v$ it is clear that the Klein factor decreases and the Lorentz factor increases, and the difference between them becomes more evident.

As in special relativity we define a parametric transformation
\begin{equation}
ct=\textrm{sin}(\phi), \hspace{1cm} x-c_0=\cos{(\phi)},    
\end{equation}
where $\phi=\arctan{(\frac{v}{c})}$. This leads  to $\gamma_k=\cos{(\phi)}$ and the Lorentz-like transformations \eqref{ktransf1} and \eqref{ktransf2} can be
seen as   an ordinary rotation by $\phi$,
that is very different from   the hyperbolic rotation which characterizes the Lorentz transformation and  is certainly not a Galilean transformation. Furthermore,  to illustrate how rich is  KST, we point out that the metric \eqref{metrica1} allows Lorentz transformations too, if we boost  in the $y-$ or $z-$direction, for instance. For a general transformation see \cite{velev2012relativistic}.

\section{Kinematics  and  consequences of Klein transformations}
Now, we  describe some physical consequences of Klein transformations, starting by kinematics. The first effect that we discuss in this spacetime, is the length dilatation. 
Let $S$ and $S'$ be two inertial frames moving with respect to each other along the $x-$direction with  relative speed $v$.
Suppose that $L_0$ is the proper length  of a horizontal rod measured by an observer in $S$. An observer in the $S'$ frame will measure a larger length, $L$, for the same rod. Both quantities, $L$ and $L_0$, are related by the equation
\begin{equation}
    L=L_0\sqrt{1+\frac{v^2}{c^2}}.
\end{equation}
This relation shows that the rod dilates,  its length being increased in the direction of the motion of $S$.

To illustrate that the time interval is also a relative quantity in this scenario, suppose that $T_0$ is the proper time measured by an observer in the $S$ frame, and that it indicates the time interval between two events. Suppose also that  $T$ is the time interval between the same two events measured by an observer in the $S'$ frame. Then, the intervals are different and are related by the expression
\begin{equation}
T=\frac{T_0}{\sqrt{1+\frac{v^2}{c^2}}},   \label{dilation}
\end{equation}
which indicates a time contraction for the $S'$ observer. In this way,
both results are the inverse of the corresponding effects of the contraction of length and the time dilatation that occur in special relativity. 

\subsection{Muon decay}
In order to better characterize  this time interval contraction effect, we analyse theoretically the muon decay. As it is well known, the muons ($\mu^{+}$ and $\mu^{-}$) are unstable particles that spontaneously decay as 
\begin{eqnarray}
\mu^{+} & \rightarrow & e^{+} + \nu_{e} + \bar{\nu}_{\mu} \nonumber \\
\mu^{-} & \rightarrow & e^{-} + \bar{\nu}_{e} + \nu_{\mu} . \nonumber
\end{eqnarray}
They have been used to demonstrate the relativistic time dilation effect (see Frisch–Smith experiment \cite{Frisch}). 

\begin{figure}[H]
    \centering
    \includegraphics[scale=0.47]{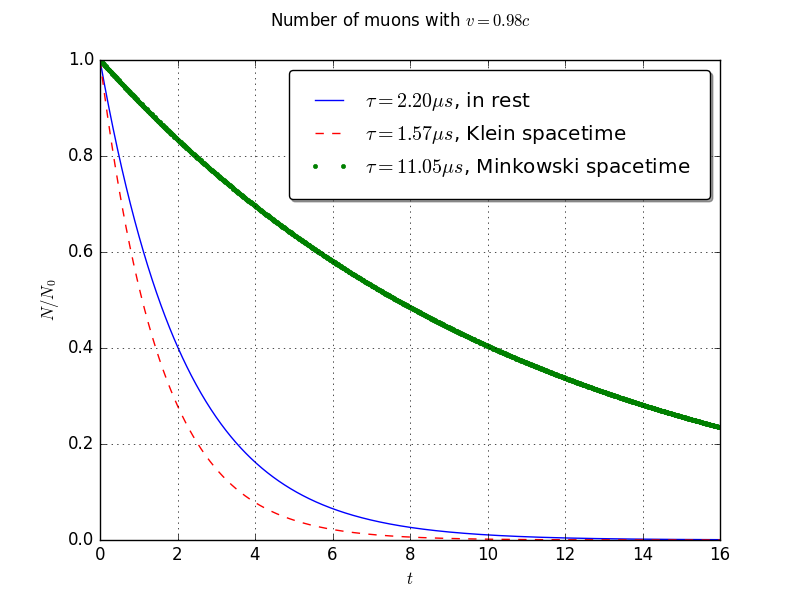}
\caption{The decay of the number of muons with time for different mean lifetimes}
    \label{fig:muondecay}
\end{figure}

The number $N$ of muons at a time $t$ is given by
\begin{equation}
    N = N_{0}\exp\left( - \frac{t}{\tau} \right),
\end{equation}
where $N_{0}$ is the number of muons at $t = 0$ and $\tau = 2.2\mu s$ is the value of the mean lifetime of the muon in rest (proper frame or co-moving frame). Supposing the muons are  moving at speed $0.98c$, and taking the frame $S$  to be attached to a laboratory on Earth and the frame $S'$  to be the rest frame of the muon, it follows from \eqref{dilation} that $\tau = 1.57 \mu s$. See  Fig. \ref{fig:muondecay}.
Therefore, in the case of Klein spacetime, the value of the mean lifetime of a moving muon measured by a frame in rest will be less than that measured by a co-moving frame (proper time). This is an interesting behavior that can be used as a probe to detect Kleinian  spacetime regions.
\subsection{Transformation of velocities}
As it could be expected, the velocity transformation is also different from that of special relativity. For one-dimensional motion, the relativistic velocity addition formula is
\begin{equation}\label{speed-transformation}
    u_x'=\frac{u_x-v}{1+vu_x/c^2},
\end{equation}
where $u_x$ is the speed of a particle in an inertial frame $S$, and $u_x'$ is the speed of the same particle in  another inertial frame, $S'$, and $v$ is the relative speed between these two frames.  With this transformation it is possible to find an observer where the light is at rest or an observer where the light speed is faster than $c$, these facts are not possible in special relativity. Although the transformation \eqref{speed-transformation} gives these strange results, the Galilean velocity composition is recovered for $v<<c$.

Observing the transverse velocity transformation,
\begin{equation}
    u'_y=\frac{u_y\sqrt{1+\frac{v^2}{c^2}}}{\left(1+\frac{vu_x}{c^2}\right)},
\end{equation}
we see that for $v<<c$, then $u_y'=u_y$ and the Galilean velocity transformation is again recovered for low velocities.
\section{Dynamical consequences}
Here  we show that the Klennian metric leads us to a new particle dynamics.
Using the  metric \eqref{metrica1}, we see that $\vec{v}\cdot \vec{v}= - v_x^2 + v_y^2 + v_z^2 $ and therefore the free particle Lagrangian becomes
\begin{equation}
   L=-m_0c^2\sqrt{1+\frac{v_x^2}{c^2}- \frac{v_y^2}{c^2}-\frac{v_z^2}{c^2}}.  \label{Lag}
\end{equation}
Now, we can find the  particle momentum $\vec{p}$. From the definition $p_i = \frac{\partial L}{\partial v_i}$, we get
\begin{gather}
    p_x=\frac{-m_0v_x}{\sqrt{1+\frac{v_x^2}{c^2}- \frac{v_y^2}{c^2}-\frac{v_z^2}{c^2}}} ,\\
    p_y=\frac{m_0v_y}{\sqrt{1+\frac{v_x^2}{c^2}- \frac{v_y^2}{c^2}-\frac{v_z^2}{c^2}}}, \\
%
    p_z=\frac{m_0v_z}{\sqrt{1+\frac{v_x^2}{c^2}- \frac{v_y^2}{c^2}-\frac{v_z^2}{c^2}}}.
\end{gather}
If we assume that the relation $p=mv$ is valid in Klein spacetime, it follows that  the rest mass is a diagonal 3D tensor $m=\text{diag}(-m_0,m_0,m_0)$, where $m_0 > 0$. The fact that the inertial response is tensorial reflects the anisotropic nature of Klein spacetime. 

\begin{figure}[H]
\begin{center}
\includegraphics[scale=0.47]{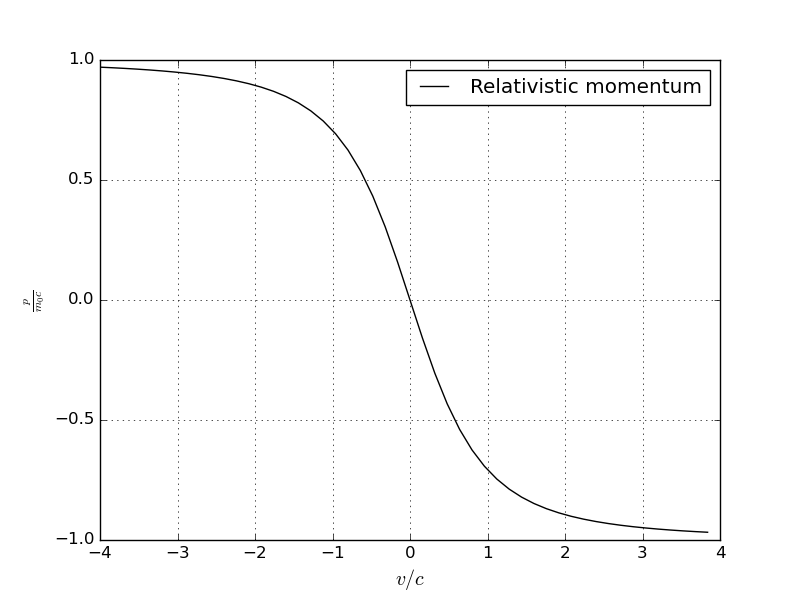}
\caption{In the figure  we see the rescaled momentum $\frac{p}{m_0 c}$ against normalized speed $v/c$. }
\label{klenianmomentum}
\end{center}
\end{figure}

If the particle is moving along the $x-$direction, then $p_x=p$, $v_x=v$, $v_y=v_z=0$, and we are left with
\begin{equation}\label{momentum}
    p=-\frac{m_0v}{\sqrt{1+\frac{v^2}{c^2}}}.
\end{equation}
From this equation we note that $p$  goes to the inverse Newtonian momentum as $v \rightarrow 0$. Conversely, it goes to $\mp m_0 c$ as $v \rightarrow \pm\infty$. In  Fig. \ref{klenianmomentum} it is possible to see the unusual behavior of the momentum, for high values of $v$.  
\FloatBarrier

Since the Lagrangian \eqref{Lag} does not depend explicitly on the coordinates, the particle momentum is conserved. From the Hamiltonian definition $H=p\dot{q}-L$, we find the relativistic energy for the particle moving in the $x$-direction,
\begin{equation}\label{eq:energy}
    E = \frac{m_0c^2}{\sqrt{1+\frac{v^2}{c^2}}}.
\end{equation}
We see from  this  equation that,  as long as the particle speeds up, the total energy approaches  zero. Also, differently from the special relativity case, where a finite rest mass particle would have an infinite momentum if $v\rightarrow \pm c$, in the present case the momentum is always finite, even  for a particle  with superluminal speed.

If we expand \eqref{eq:energy}, for small values of  $v/c$, we find the relation
\begin{equation}
  E\approx m_0c^2-\frac{m_0v^2}{2}, \label{Eexp}
\end{equation}
where we can identify the kinetic energy as $T={m_0v^2}/{2}$ and the rest energy by $m_0c^2$. 
This is very different from the special relativity case, where both energies are added but reinforces the anisotropic character  of the inertial mass.  Note that, for motion along the $y$ or $z$ directions Eq. \eqref{Eexp} comes with the usual $+$ sign in front of the kinetic energy term.

For a general motion, it is easily deduced that the energy becomes 
\begin{equation}
    E=\frac{m_0c^2}{\sqrt{1+v_x^2-v_y^2-v_z^2}},
\end{equation}
as described by \cite{velev2012relativistic}.

\begin{figure}[H]
\begin{center}
	\includegraphics[scale=0.47]{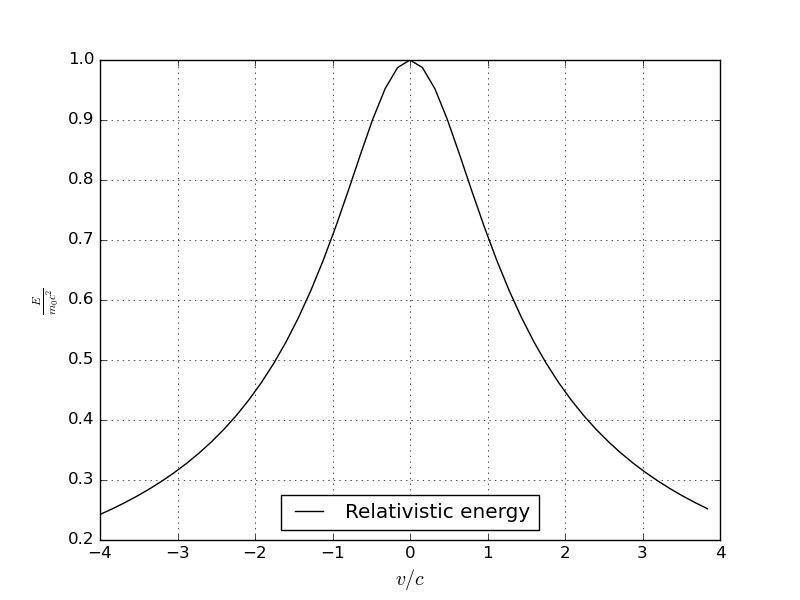} \quad
	\caption{The figure shows the  point particle  rescaled energy  $E/m_0c^2$ against the normalized speed $v/c$. } \label{gdimotes}
\end{center}
\end{figure}

Taking  the four-momentum as $p^{\mu}=(E/c,p,0,0)$ the conservation equation, $p^{\mu}p_{\mu}=m_0^2c^2$,  has the explicit form
\begin{equation}
    E^2+p^2c^2=m_0^2c^4, \label{E2}
\end{equation}
if the motion happens  along the  $x-$direction.
It is very important to note that this relation corresponds to an exotic conservation equation, that was previously discussed in \cite{wang2016modern}, who has also considered negative masses. The energy behavior \eqref{eq:energy} can be seen in the Fig. \ref{gdimotes}.
\FloatBarrier

\section{ Unidimensional motion in an uniform Newtonian field}

As an application of the new mechanical ideas developed above,  we present a single particle motion due to a uniform Newtonian gravitational field. Using a more geometrical approach, let us consider a particle moving along some world line $x^{\mu}(\lambda)$  under a uniform Newtonian gravitational potential $V(r)$. The associated Lagrangian is
\begin{equation}
    L=-m_0c\sqrt{\eta_{\alpha\beta}\frac{dx^{\alpha}}{d\lambda}\frac{dx^{\beta}}{d\lambda}}-V(r).
\end{equation}
It is clear that the above Lagrangian is not covariant, but for our proposes this fact is not important. With $x^{\mu}=(t,x)$, the Lagrangian provides the following Euler-Lagrange equations of motion for the particle
\begin{gather}
    \frac{d}{d\lambda}\left(\frac{\dot{t}}{\sqrt{\eta_{\alpha\beta}\dot{x}^{\alpha}\dot{x}^{\beta} }}\right)=0, \label{eq:rindler-tempo}\\
    m_0c^2\frac{d}{d\lambda}\left(\frac{\dot{x}}{\sqrt{\eta_{\alpha\beta}\dot{x}^{\alpha}\dot{x}^{\beta} }}\right)=-\frac{\partial V(r)}{\partial r}.\label{eq:Movimento-ciruclar1}
\end{gather}
Since we want to investigate the motion of a particle under a constant gravitational field, we choose $V=m_0g x$, and $\lambda=t$, then   equation \eqref{eq:Movimento-ciruclar1}
gets the following form
\begin{equation}\label{eq:rindler1}
    \frac{d}{dt} \left( \frac{v}{\sqrt{1+\frac{v^2}{c^2}}} \right) = g.
\end{equation}
In fact, this equation describes the motion of a particle under constant acceleration seen by an observer in a rest frame $S$, that implies in the solution
\begin{equation}
    x=k_1\pm\sqrt{\frac{c^4}{g^2}-c^2\left(t-k_2\right)^2},
\end{equation}
where $k_1$ and $k_2$ are integration constants. This equation can be easily cast in the form
\begin{equation}\label{eq:elipse}
(x-k_1)^2+c^2(t-k_2)^2=\frac{c^4}{g^2},    
\end{equation}
which describes an ellipse in the $t-x$ plane.
For simplicity, we admit $k_2=0$, then the velocity of the particle can be expressed as
\begin{equation}
   v = \mp\frac{c^2t}{\sqrt{\frac{c^4}{g^2}-c^2t^2}}.
\end{equation}
Observing this equation, we see a singular behavior for $t=\pm\frac{c^2}{g}$. This fact leads us to restrict the problem to the time interval $(-\frac{c^2}{g}, +\frac{c^2}{g})$. We see that for higher values of $g$, this time interval becomes smaller.

\begin{figure}[h]
     \centering
    \includegraphics[scale = 0.47]{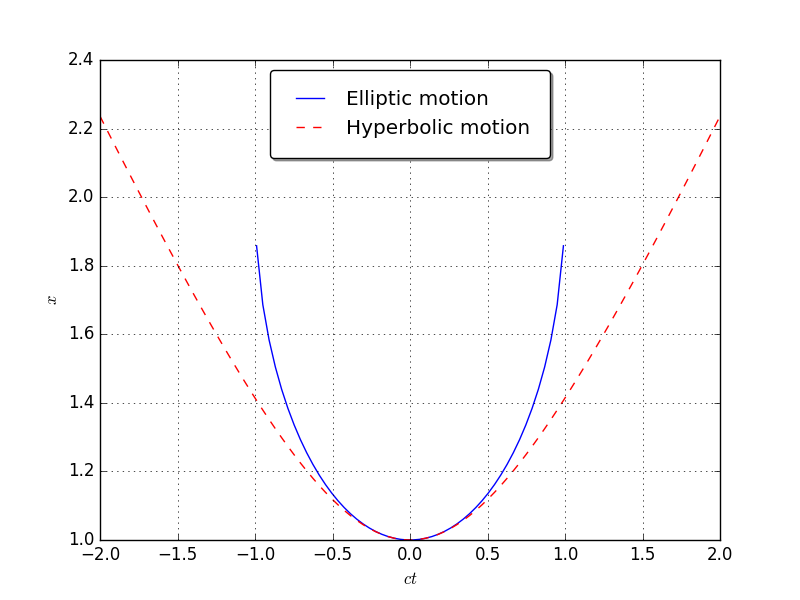}
    \caption{Considering $c=g=1$,$k_1=1$ and $k_2=0$, we show a comparison  between the  uniformly accelerated motion in the $x-$direction of Klein spacetime   and in the corresponding direction in Minkowski spacetime.}
    \label{rindler-like}
\end{figure}

\begin{figure}
    \centering
    \includegraphics[scale = 0.47]{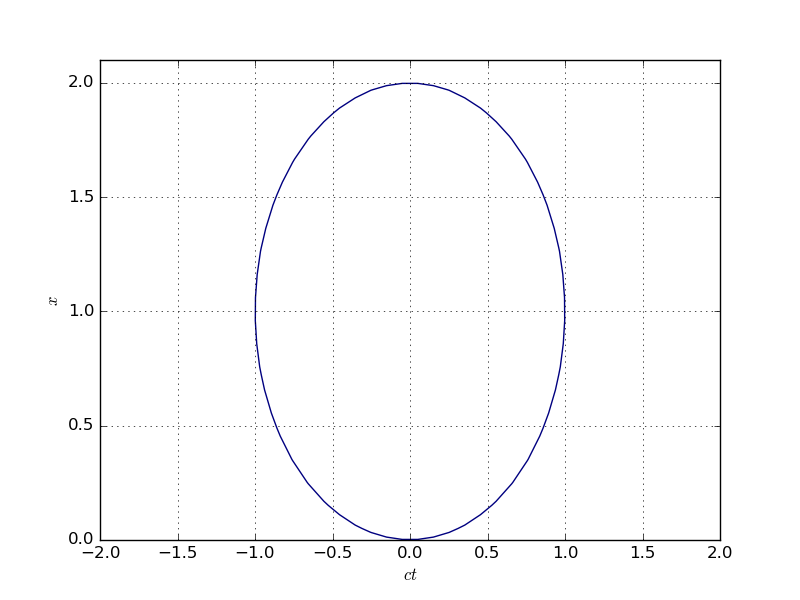}
    \caption{Considering $c=g=1$, $k_1=1$ and $k_2=0$, we plot the complete elliptic ``trajectory'' according to \eqref{eq:elipse}.}
    \label{fig:eq26}
\end{figure}

Figure \ref{rindler-like} illustrates the particle motion under constant acceleration as expected from special relativity and from our case (elliptic motion). We can see, for  $t\rightarrow \pm\frac{c^2}{g}$, that the particle seems to disappear. However, if we plot the full curve we can interpret the ``forbidden'' regions as  the particle traveling back in time (antiparticle) as it is the case of tachyons moving in Minkowski spacetime. Accordingly, at any given instant of time, particle and antiparticle appear at opposing sides of the ellipse shown in Fig. \ref{fig:eq26}.

This example is an illustration of how KST can give rise to physical issues as causality, unbound velocities, closed time-like curves and so on. Nevertheless, from the geometrical point of view, this space-time carries interesting features.
\FloatBarrier
\section{Some quantum consequences}
It is well known that the principles of Einstein's Special Relativity gave rise to new quantum mechanical models. The Klein-Gordon equation, for example, is a simple generalization of Schr\"odinger quantum mechanics in the framework of Special Relativity. Dirac, however, proposed an equation that has  fundamental importance in the understanding of the relativistic spinning particle dynamics \cite{Dirac1928}. In this section, we dedicate a few words about the Dirac equation in Klein spacetime. First, we revisit some ideas already discussed  in this direction and then we develop complementary new ones.  

Following Alty's results \cite{alty1994kleinian}, we adopt a prescription for Kleinian gamma matrices,$\gamma^{\mu}_{K}$, as 
\begin{equation}
\gamma^{0}_{K}=
\begin{pmatrix}
I & 0 \\
0 & -I 
\end{pmatrix},
\end{equation}

\begin{equation}
    \gamma^{1}_{K}=
\begin{pmatrix}
0 & i\sigma_{1} \\
-i\sigma_{1} & 0 
\end{pmatrix},
\end{equation}
\begin{equation}
    \gamma^{i}_{K}=
\begin{pmatrix}
0 & \sigma^{i} \\
\sigma^{i} & 0
\end{pmatrix}, 
\end{equation}
where $I$ is the identity matrix. With this choice the relation between the Kleinian gamma matrices and the spacetime metric remains  the usual,  
\begin{equation*}
    \{\gamma^{\mu}_{K},\gamma^{\nu}_{K}\}=2\eta^{\mu\nu}I.
\end{equation*}
We note that,  for $\mu=0,2,3$, the gamma matrices $\gamma^{\alpha}_{K}$ are in the Pauli-Dirac representation.  However,  $\gamma^{1}_K$ is not. This last one is related to the usual gamma prescription in Lorentz spacetime, $\gamma^{1}_{L}$, by the transformation  $\gamma^1_{K}=i\gamma^{1}_{L}$.

\subsubsection{Free particles}
Now that we have pointed out  these settings, we  analyse the free spin-$\frac{1}{2}$ particle motion. In this spacetime the Dirac equation gets the form
\begin{equation}
    \left( i\hbar\gamma^{\mu}_{K} \partial_{\mu}-m_0c\right)\psi=0,
\end{equation}
 where $x^0=ct$. For this case, we  assume that the particle has the plane wave  profile
 \begin{equation}
     \psi = N e^{i(\vec{p}\cdot\vec{r} -Et)} \begin{bmatrix}
       \phi(p)\\
        \chi(p)
        \end{bmatrix},
\end{equation}
where $N$ is the normalization factor,  $\phi(p)$ and $\chi(p)$ are the spinor components. In terms of the Pauli matrices, $\vec{\sigma}$, the Dirac equation  becomes
\begin{equation}
    \begin{bmatrix}
       E-m_0c  & -\vec{p}\cdot\vec{\sigma} \\
        \vec{p}\cdot\vec{\sigma}  &    -E-m_0c
        \end{bmatrix} \begin{bmatrix}
       \phi(p)\\
        \chi(p)
        \end{bmatrix} = 0 .
\end{equation}
Here we are concerned with the one-dimensional motion along the $x-$direction,    such that $p_y=p_z=0$, that leads us to  the following equation
\begin{equation}
     \begin{bmatrix}
       E/c-m_0c^2  & -ip_x\sigma_1 \\
        ip_x\sigma^x  &    -E/c-m_0c^2
        \end{bmatrix} \begin{bmatrix}
       \phi(p)\\
        \chi(p)
        \end{bmatrix} = 0.
 \end{equation}
The non-trivial solutions  of the above system occur if the matrix determinant is zero. Applying this condition we again arrive at the dispersion relation \eqref{E2}
\begin{equation}\label{eq:spectrum}
 E^2=m_0^2c^4-p^2c^2,    
\end{equation}{}
where $p=p_x$.

As usual, we have two solutions for negative energy states expressed as
\begin{equation}
  \psi^{\textrm{up}}_{(-)}= N_{(-)}e^{i(\vec{p}\cdot\vec{x} -E^{(-)}t)} \begin{bmatrix}
                0\\
                \frac{-ip_x c}{E^{(-)}-m_0c^2}\\
                1\\
                0
                \end{bmatrix},
                \end{equation}
 \begin{equation}
 \psi^{\textrm{down}}_{(-)}= N_{(-)}e^{i(\vec{p}\cdot\vec{x} -E^{(-)}t)} \begin{bmatrix}
                \frac{-ip_xc}{E^{(-)}-m_0c^2}\\
                0\\
                1\\
                0
\end{bmatrix},
\end{equation}
where $\psi^{\textrm{up}}_{(-)}$ represents the spin up particle, and $\psi^{\textrm{down}}_{(-)}$ represents the spin down particle and we have defined
\begin{equation*}
    N_{(-)} \doteq \sqrt{\frac{m_0c-E^{(-)}/c}{2Vm_0c}.}
\end{equation*}
Similarly, we also have two positive energy state solutions 
\begin{equation}
  \psi^{\textrm{up}}_{(+)}= N_{(+)}e^{i(\vec{p}\cdot\vec{x} -E^{(+)}t)} \begin{bmatrix}
                1\\
                0\\
                0\\
                \frac{ip_x c}{E^{(+)}+m_0c^2}\\
                \end{bmatrix},
 \end{equation}
 
 \begin{equation}
 \psi^{\textrm{down}}_{(+)}= N_{(+)}e^{i(\vec{p}\cdot\vec{x} - E^{(+)}t)} \begin{bmatrix}
                0\\
                1\\
                \frac{ip_xc}{E^{(+)}+m_0c^2}\\
                0
\end{bmatrix},
\end{equation}
and we have defined
\begin{equation*}
   N_{(+)} \doteq \sqrt{\frac{m_0c+E^{(+)}/c}{2Vm_0c}}.
\end{equation*}
\subsubsection{Forbidden states}
\begin{figure}
    \centering
    \includegraphics[scale=0.45]{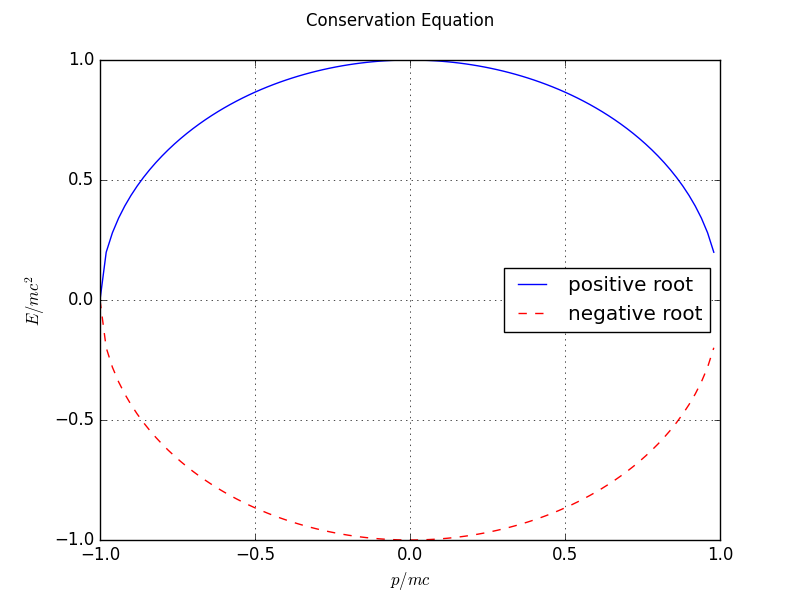}
    \caption{Dispersion relation for a free fermion moving in the $x$-direction, as given by \eqref{eq:spectrum}, showing a single continuous energy band restricted to the interval $-m_0c^2 <E< m_0 c^2$.}
     \label{fig:conservation-eq} 
\end{figure}

\begin{figure}
    \centering
    \fbox{\includegraphics[scale=0.55]{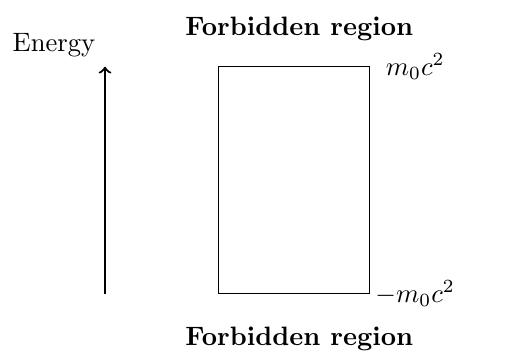}}
    \caption{Single energy band of the free unidimensional Dirac particle moving along $x-$direction in Klein spacetime.}
     \label{fig:band-energy} 
\end{figure}

Looking at the free particle dispersion relation \eqref{eq:spectrum}, we infer that there is just one band of allowed energy levels, which lays between $-m_0 c^2$ and $m_0 c^2$, as illustrated in Fig. \ref{fig:conservation-eq}.
\FloatBarrier
In Fig. \ref{fig:band-energy} is another representation of the energy spectrum showing again the single energy band shown in Fig. \ref{fig:conservation-eq}. This is the complement of the usual Dirac spectrum of the free particle.
\FloatBarrier
Because of this gapless energy structure there is no minimum energy requirement to promote a fermion from a negative  energy state to a positive one, but there are lower and upper bounds for the energy.  As a consequence, there is a limit in the amount of energy that a particle can either absorb or emit, quite differently from the usual relativistic case. We also note that due to these  energy limits, the catastrophic electronic transition of   relativistic quantum mechanics \cite{ryder1996quantum} is prevented. 

\begin{figure}[H]
    \centering
    \includegraphics[scale=0.450]{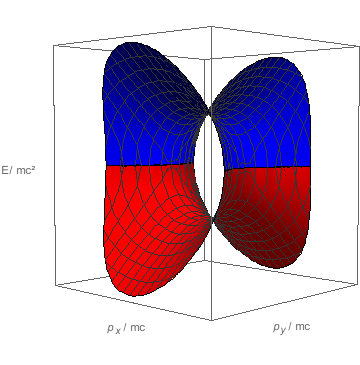}
    \includegraphics[scale=0.450]{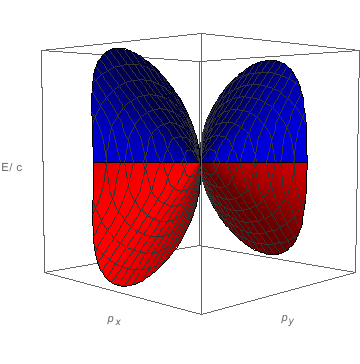}
    
    \caption{Two-dimensional energy surface representing the dispersion relation\eqref{eq:hiperboloide}. The blue portion of the surface represents the positive root and the red one the negative root of the dispersion relation.} {The second figure, the elliptical cone, corresponds to  Eq. \eqref{eq:conseq5}.}
     \label{fig:conservation-eq2} 
\end{figure}

\begin{figure}[H]
    \centering
    \includegraphics[scale=0.450]{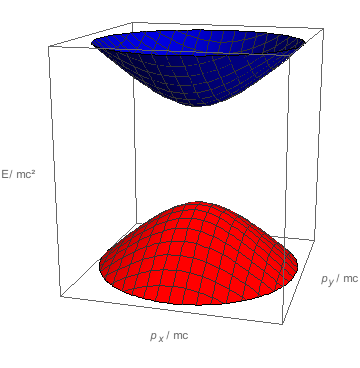}
    \includegraphics[scale=0.450]{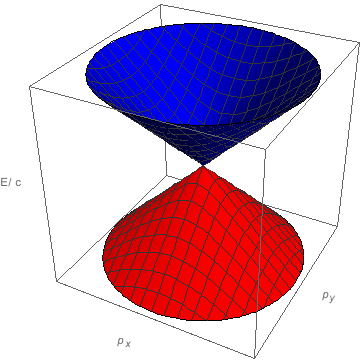}
    
    \caption{Two-dimensional energy surface representing the dispersion relations \eqref{eq:elliptic} and \eqref{eq:conseqx}. The blue portion of the surface represents the positive root and the red one the negative root of the dispersion relation.}
     \label{fig:conservation-eq4} 
\end{figure}

By considering the motion in the $x-y$ plane, the dispersion relation becomes 
\begin{equation}\label{eq:hiperboloide}
    \frac{E}{m_0c^2} = \pm\sqrt{1 - \left(\frac{p_{x}}{m_0c}\right)^2 + \left(\frac{p_{y}}{m_0c}\right)^2},
\end{equation}
a hyperboloid, as depicted in Fig. \ref{fig:conservation-eq2}. This type of behavior is characteristic of hyperbolic metamaterials \cite{poddubny2013hyperbolic}.
From this hyperboloid, we see that there is no  $p_x-p_y$ rotational symmetry. Also note that by taking any $p_y$,  the energy band increases its bounds.   If we consider the photons,  for example, the dispersion relation changes and gets the form
\begin{equation}\label{eq:conseq5}
    \frac{E}{c} = \pm\sqrt{p_{y}^2  -p_{x}^2},
\end{equation}
valid or $p_y\geq p_x$. The corresponding  diagrams are represented in Fig. \ref{fig:conservation-eq2}, and for a comparison  we represent the usual dispersion relations in Minkowski space-time, Eqs. \eqref{eq:elliptic} and \eqref{eq:conseqx}, in  Fig. \ref{fig:conservation-eq4}.
\FloatBarrier
\begin{equation}\label{eq:elliptic}
    \frac{E}{m_0c^2} = \pm\sqrt{1 + \left(\frac{p_{x}}{m_0c}\right)^2 + \left(\frac{p_{y}}{m_0c}\right)^2},
\end{equation}
\begin{equation}\label{eq:conseqx}
    \frac{E}{c} = \pm\sqrt{p_{y}^2  +p_{x}^2},
\end{equation}
\FloatBarrier

\subsubsection{Weyl fermions and helicity}
Now we look at  a particular case where we make $m_0=0$ in the Dirac equation, the so-called Weyl spinor case, that was related with the neutrino  theoretical description \cite{greiner2000relativistic}. The motion of Weyl fermions in Kleinian spacetime is severely restricted by Eq.  \eqref{eq:hiperboloide} which limits the $x-$component of the momentum. It is also clear that there cannot be Weyl fermions moving only in the $x-$direction.

Still thinking only in x-diretion movements, all other one-half spinning particles moving in this direction have a frame-dependent helicity, $h$. Furthermore, a right-handed massive particle, $h>0$, has its spin in the same direction of its momentum $p$ but in the opposite direction of its motion. Alternatively, a massive left-handed particle, $h<0$, has its spin in the same direction of its motion but in the opposite direction of its momentum. These facts are consequences of the unusual  {tensorial mass character}.

\section{Analogue Kleinian fermions}
The dispersion relation in  Kleinian relativity can be realized using the  analogue Dirac equation designed for metamaterials, developed in \cite{tan2014photonic,ahrens2015simulation}. To apply their ideas here, we start this section with the following Maxwell equations,
\begin{gather}
    \vec{\nabla} \times \vec{E}=\mu_r\mu_0\frac{\partial \vec{H}}{\partial t}\\
    \vec{\nabla} \times \vec{H}=-\epsilon_r\epsilon_0\frac{\partial \vec{E}}{\partial t},
\end{gather}
that describe electromagnetic waves in one-dimensional metamaterials. Assuming that the wave propagates in the $x-$direction, the equations can be simplified and written in the form
\begin{gather}
    \partial_x E_z=i\omega\mu_r\mu_0H_y,\\
    \partial_xH_y=-i\omega \epsilon_r\epsilon_0E_z,
\end{gather}
where $\epsilon_r$  and $\mu_r$  are the constitutive parameters. We define the analogue fermion in Kleinian relativity $\psi$ in terms of the electromagnetic wave component through the relation 
\begin{equation}
    \psi=
        \begin{pmatrix}
            \sqrt{\epsilon_0}E_z\\
            \sqrt{\mu_0}H_y
        \end{pmatrix}.
\end{equation}
The wave propagation equation may  be stated as
\begin{equation}
\left[-i\sigma_x \partial_x+\sigma_z m(\omega)\right]\psi=E(\omega)\psi.
\end{equation}
This is the one-dimensional Dirac equation in natural units($c=\hbar=1$). The effective energy, $E(\omega)$, and effective mass $m(\omega)$, and the wave number $k$ are functions of the constitutive parameters in the following form
\begin{gather}
 m(\omega)=\frac{\omega\sqrt{\mu_0\epsilon_0}}{2}\left[\epsilon_r(\omega)-\mu_r(\omega) \right], \\
E(\omega)=-\frac{\omega \sqrt{\mu_0\epsilon_0}}{2}\left[\epsilon_r(\omega)+\mu_r(\omega)\right],\\
k^2=\omega^2\epsilon_0\epsilon_r(\omega)\mu_0\mu_r(\omega).\label{eq:K}
\end{gather}
From these equations it is straightforward to show that the dispersion relation is
\begin{equation}
    E(\omega)^2=m(\omega)^2+ k^2.
\end{equation}
\subsection{Mimicking the Kleinian fermions}
Here we assume, for our proposes, that the relative permeability and the relative permittivity have opposite signs, this is $\epsilon_r(\omega)<0$, $\mu_r(\omega)>0$ (or the other case $\epsilon_r(\omega)>0$, $\mu_r(\omega)<0$). { As a result from \eqref{eq:K}, we have the following dispersion relation
\begin{equation}\label{eq:analog-energy}
    E(\omega)^2=-k^2 +m(\omega)^2,
\end{equation}
that  characterize the Kleinnian  conservation equation, where the effective parameters are}
\begin{gather}
 m(\omega)=\frac{\omega\sqrt{\mu_0\epsilon_0}}{2}\left[\epsilon_r(\omega)+|\mu_r(\omega)| \right] \label{eq:massa-ef1},\\
    E(\omega)=-\frac{\omega\sqrt{\mu_0\epsilon_0}}{2}\left[\epsilon_r(\omega)-|\mu_r(\omega)|\right]\label{eq:energia-ef1}.
\end{gather}
From equation $\eqref{eq:massa-ef1}$, no negative  or null mass analogue particles are allowed, then no analogue Weyl fermions must move in this direction. In addition, for this case, negative energy states are reproduced when $\epsilon_r(\omega)>|\mu_r(\omega)|$, while the positive energy states are simulated when $\epsilon_r(\omega)<|\mu_r(\omega)|$.

In the opposite way, if we assume that $\epsilon_r(\omega)<0$, $\mu_r(\omega)>0$, the relation \eqref{eq:analog-energy} still holds, then the analogue mass and energy expressions are written as
\begin{gather}
 m(\omega)=-\frac{\omega\sqrt{\mu_0\epsilon_0}}{2}\left[|\epsilon_r(\omega)|+\mu_r(\omega) \right], \label{eq:massa-ef2}\\
E(\omega)=-\frac{\omega\sqrt{\mu_0\epsilon_0}}{2}\left[\mu_r(\omega)-|\epsilon_r(\omega)|\right].\label{eq:energia-ef2}
\end{gather}
Both cases are interesting since they provide different signs for $m_0$, which could describe the motion of a fermion along the $x-$direction in Klein spacetime.    
 However, there are some limitations in this analogue model as stated by \cite{ahrens2015simulation}.  This adaptation  indicates a way to simulate the Klein-Minkowski metric signature transition using fermions as test particles, complementing the discussion of \cite{figueiredo2016modeling}.

\section{Final Remarks}
In this work, we developed a relativistic mechanical model for  single point particles in Klein spacetime. A summary of some of our results and   a comparison between known mechanical models is presented in Table I.
\begin{table}[H]
\begin{flushleft}
\begin{tabular}{|c|c|c|c|c|}
\hline
\textbf{Model}  & $v_x$ & $p_x$ &   $E$ \\
\hline                    
 Minkowskian & $|v_x| < c$ &  $|p_x| < \infty$ &    unbound\\
 \hline
\textsc{Kleinian} &  $0<|v_x| < \infty$     &  $|p_x| < m_0c$  & bound \\
 \hline
 Tachyonic &  $c<|v_x| < \infty$   & $ m_0c<|p_x| < \infty$  & unbound\\
 \hline
 Newtonian & $0<|v_x|<\infty$& $0<|p_x| < \infty$  & unbound \\
   \hline
\end{tabular}
\end{flushleft}
\label{tab1}
\caption{Single particle mechanical properties in different  models {for  one-dimensional movement in the $x-$direction.} }
\end{table}
From this table we see that the Kleinian  free particles  do not have an upper or lower speed limit, but their momentum is bound. This is in accordance with the fact that these particles have a limited range of energy values, and this is completely different from the tachyon and special relativity description of particles.
However, free tachyons and free Klein particles 
moving in the $x-$direction share some common features as the possibility to reach superluminal speeds and as the fact that they could have zero total energy and still carry a momentum $m_0c$. 
Furthermore, it is also important to stress that:
\begin{enumerate}
    \item The existence of the so-called exotic matter, as stated by Wang \cite{wang2016modern}, seems to be a consequence of Klein spacetime, since it allows a negative  mass component. 
     \item The first postulate of Einstein's Special Relativity does not agree with this new scenario, only the second postulate is in agreement with it. This is a clear consequence of the non-Minkowskian geometry.
     \item If, somehow, the time contraction by muon decay were measured then it should be an evidence of the existence of a patch of KST somewhere.
     \item  There is an upper bound for
     the $x-$direction momentum of a Weyl fermion in Kleinian spacetime, even if the particle also  moves in other directions, as suggested by Eq. \eqref{fig:conservation-eq2}.
     \item Metamaterials can only mimic some  proprieties of the KST, and we should not expected that all the metric derived effects can be reproduced in laboratories with this technology. Then, part of our results only happens in a true, non-effective, Klein geometry, that have their importance  in theoretical grounds.
\end{enumerate}

Perhaps the main contribution of this article is to complement  Alty's discussion \cite{alty1994kleinian} on Kleinian spacetime effects, adding some relativistic analyses to it. Furthermore, we expect that our developments will inspire new analogue models for Klein spacetime, and new investigations, such as in the quantum field theory realm, in special, in order to clarify the unusual Dirac sea picture. In addition, we hope that our results constitute an important step in the development of this curious and pathological scenario, which we call \textit{Kleinian Relativity}.
\acknowledgments{This work was supported by Conselho Nacional de Desenvolvimento Cient\'{i}fico e Tecnol\'{o}gico (CNPq), grants Nos. 152124/2019-5 (ABB) and 307687/2017-1 (FM). We also thank the referee for its meaningful contributions.}
\bibliography{refs}
\end{document}